\documentclass[aps,prl,reprint,altaffilletter,superscriptaddress]{revtex4-1}
\usepackage[utf8]{inputenc}

\usepackage{braket}
\usepackage{amsmath}
\usepackage{graphicx}
\usepackage{color}
\usepackage{physics}
\usepackage{siunitx}

\begin{document}
\author{Sudipta Kundu}
\altaffiliation[Present address: ]{Department of Materials Science and Engineering, Stanford University, Stanford, CA, USA}
\affiliation{Centre for Condensed Matter Theory, Department of Physics, Indian Institute of Science, Bangalore 560012, India}
\author{Tomer Amit}
\affiliation{Department of Molecular
Chemistry and Materials Science, Weizmann Institute of
Science, Rehovot 7610001, Israel}
\author{H. R. Krishnamurthy}
\affiliation{Centre for Condensed Matter Theory, Department of Physics, Indian Institute of Science, Bangalore 560012, India}
\author{Manish Jain$^*$}
\affiliation{Centre for Condensed Matter Theory, Department of Physics, Indian Institute of Science, Bangalore 560012, India}
\author{Sivan Refaely-Abramson$^*$}
\affiliation{Department of Molecular
Chemistry and Materials Science, Weizmann Institute of
Science, Rehovot 7610001, Israel}

\title{Exciton fine structure in twisted transition metal dichalcogenide heterostructures}

\begin{abstract}
    Moir\'{e} superlattices of transition metal dichalcogenide (TMD) heterostructures give rise to rich excitonic phenomena associated with the interlayer twist angle and induced changes in the involved quantum states.
    Theoretical calculations of excitons in such systems are typically based on model moir\'{e} potentials to mitigate the computational cost. However, an \textit{ab initio} understanding of the electron-hole coupling dominating the excitations is crucial to realize the twist-induced modifications of the optical selection rules. In this work we use many-body perturbation theory to compute and analyze the relation between twist angle and exciton properties in twisted TMD heterostructures. We present a general approach for unfolding excitonic states from the moir\'{e} Brillouin zone onto the Brillouin zones of the separate layers. Applying this method to a twisted MoS$_2$/MoSe$_2$ bilayer, we find that the optical excitation spectrum is dominated by mixed transitions between electrons and holes with different momenta in the separate monolayers, leading to unexpected and angle-dependent hybridization between interlayer and intralayer excitons. Our findings offer a design pathway for tuning exciton layer-localization in TMD heterostructures as a function of twist angle.
\end{abstract}

\maketitle

Moir\'{e} patterns generated due to a lattice mismatch between layers of two-dimensional materials serve as an emerging platform for novel correlated electronic and excitonic physics.
In particular, twisted heterostructures of transition metal dichalcogenides (TMDs)  with a type-II band alignment exhibit intriguing optical properties due to the varying exciton localization associated with the twist-induced moir\'{e} potential~\cite{ttmd1,ttmd2,ttmd3, tran2019evidence, Seyler2019, moire1}. 
These excitons, broadly explored experimentally and theoretically~\cite{jin2018ultrafast, ttmd7,ttmd9,ttmd12, karni2019infrared, tran2019evidence, karni2022structure,jin2019moire}, are shown to exhibit both interlayer and intralayer nature which depends on the underlying quantum-state modifications stemming from the sublattice composition~\cite{rivera2018interlayer, Yu2015, kumar2021spin}. The twist-induced excitations can be detected through the exciton fine structure in optical absorption and emission measurements~\cite{regan2022emerging, barre2022optical}, pointing to structural tunability of the exciton decay mechanisms and  lifetimes~\cite{Seyler2019,  ttmd6, rivera2016valley,  Jauregui2019, Choi2021}.

The interlayer twist angle dictates the relation between the moir\'{e} Brillouin Zone (MBZ) and the unit-cell Brillouin Zones (UBZs) of the separate layers, inducing an associated moir\'{e} potential~\cite{thry1, ruiz2019interlayer}. This leads to optically-allowed electron-hole transitions between states of different momenta in the UBZs, which can be associated with the exciton fine structure in absorption~\cite{Brem2020, Seyler2019, kunstmann2018momentum, wu2017topological}. From an electronic structure perspective, the moir\'{e} potential also introduces structural effects which determine the momentum and spin selection rules responsible for the optical excitations. Atomic reconstruction is induced through interlayer mismatch and generates non-uniform strain~\cite{maity2021reconstruct,carr2018reconstruct,enaldiev2020reconstruct, kang2013electronic}, changing the atomic structure associated with the original layers with dependence on the size of the moir\'{e} periodicity. Together with the interlayer coupling and dielectric screening, these effects modify the electron and hole localization and the respective bandstructures~\cite{bradley2015probing, slobodkin2020quantum, ttmd3,kundu1,moire1}.

While theoretical assessments of excitons in moir\'{e} heterostructures can be achieved through effective Hamiltonians~\cite{thry1, wu2017topological, thry4}, a comprehensive understanding of the relation between these twist-induced structural modifications and the exciton fine structure demands a predictive approach.  Density functional theory (DFT) can be used to capture the nature of the atomically-reconstructed electronic wavefunctions~\cite{thry3, maity2021reconstruct}. However, a structure-sensitive excitonic description requires a first-principles assessment of the dielectric function and the electron-hole coupling. These can be achieved through many-body perturbation theory within the GW and Bethe-Salpeter equation (GW-BSE) approximation~\cite{gw1,gw2,bse2,bse}. 
Yet, providing reliable GW-BSE computations of large moir\'{e} cells is extremely challenging, thus they have been applied primarily through interpolation of commensurate bilayers~\cite{tran2019evidence, karni2022structure, barre2022optical}, or by coupling the moir\'{e} electronic wavefunctions to specifically explore intralayer states~\cite{Naik2022intralayer}. However, while these approaches work well for small twist angles, they do not supply a general \textit{ab initio} understanding of the relation between twist angle and the exciton fine structure. 

In this letter, we present a new approach to study the effect of twist angle on the exciton nature and optical selection rules in TMD hetero-bilayers using GW-BSE. We develop a scheme for unfolding the electronic bandstructure  and exciton components onto the UBZs of the constituent layers,  and demonstrate it on a twisted MoS$_2$/MoSe$_2$ heterostructure with a relative rotation of 16\si{\degree}. This twist angle allows transitions between distinct high-symmetry points in the UBZs, determined by the atomic reconstruction and interlayer mismatch. Our analysis reveals a unique momentum-mixed excitonic nature, with states that are comprised of both inter- and intra-layer electron-hole excitations. Our findings suggest  a direct relation between the observed exciton spectral features and the underlying structural changes in twisted TMD heterostructures, offering tunable excitonic properties upon the choice of interlayer twist angle.  

\begin{figure}
    \centering
    \includegraphics[scale=0.35]{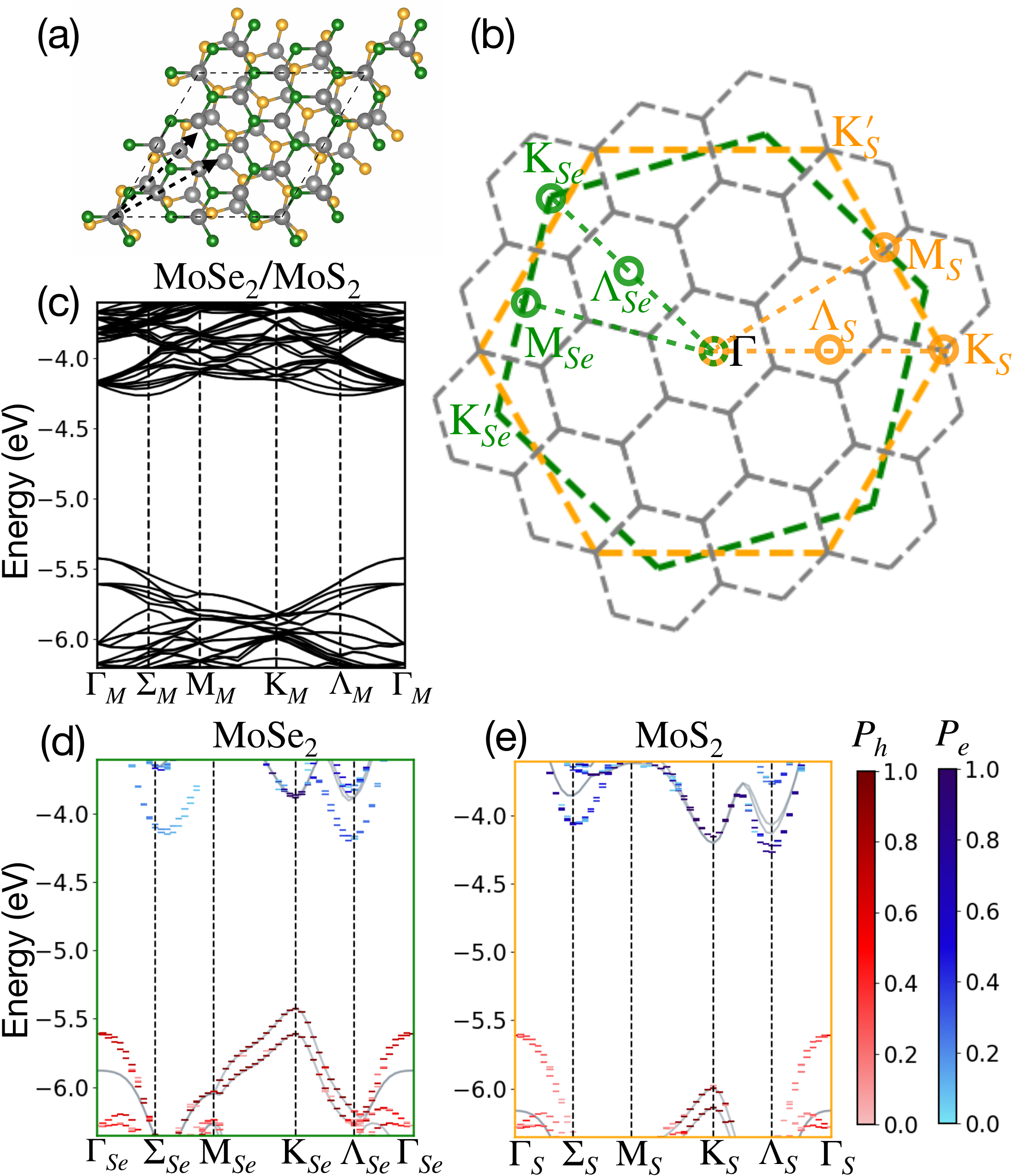}
    \caption[DFT band structure and unfolded band structures in UBZ]{(a): Structure of a 16\si{\degree} twisted MoS$_2$/MoSe$_2$ heterostructure (b): Grey hexagons represent the MBZ. 
    The green and orange hexagons correspond to the UBZs of MoSe$_2$ and MoS$_2$ respectively. The high symmetry points in the UBZs are shown. (c): DFT band structure of the heterostructure in the MBZ. The vacuum is set to zero. 
    (d) and (e): The DFT band structure unfolded from the MBZ
    to the UBZ of MoSe$_2$ and MoS$_2$ respectively. The bandstructures of the respective monolayers are shown as solid grey lines. }
    \label{fig:fig1}
\end{figure}

The examined MoS$_2$/MoSe$_2$ heterostructure is schematically shown in Fig.~\ref{fig:fig1}(a). We focus on a 16\si{\degree} twisted heterostructure as a computationally-tractable, yet useful example for studying the associated optical properties using GW-BSE.
The moir\'{e} superlattice consists of 13 unit cells of MoS$_2$ and 12 unit cells of MoSe$_2$, with a moir\'{e} length of 11.25 \AA{}. We use DFT to relax the atomic structure while including the effect of the twist-induced atomic reconstruction (see SI for the computational details). This is crucial for properly accounting for the electron-hole coupling that determines the excitonic states.  
Fig.~\ref{fig:fig1}(b) shows the MBZ (grey hexagons), as well as the UBZs of MoSe$_2$ and MoS$_2$ (green and
orange hexagons, respectively).  The $\Gamma_{Se}$, $\textrm{M}_{Se}$, $\Lambda_{Se}$, $\textrm{K}_{Se}$ and K$^{\prime}_{Se}$ points of the MoSe$_2$ UBZ and the $\Gamma_{S}$ of the MoS$_2$ UBZ fold onto the $\Gamma_M$ point of the MBZ. In contrast, the $\textrm{K}_S$ and K$^{\prime}_S$ points of the MoS$_2$ UBZ fold onto the K$_M$ and K$^{\prime}_M$ points of the MBZ, respectively; $\textrm{M}_{S}$ folds onto $\textrm{M}_{M}$, and $\Lambda_{S}$ fold onto a point nearby the $\Lambda_{M}$ of the MBZ.  
As an important outcome, for this chosen twist angle, there cannot be any optically-direct K$_{Se}$-K$_{S}$ interlayer exciton transitions. On the other hand, the coupling between hole states around K$_{Se}$ or $\Gamma$ to electrons around $\Lambda_{S}$ becomes available.
To understand the optical transitions allowed for such a heterostructure composition, we present a scheme for unfolding excitons. 

As a first step, we unfold the electronic bands composing the BSE excitons, to realize their layer contributions and their components within the UBZ. 
Fig.~\ref{fig:fig1}(c) shows the calculated DFT bandstructure of the examined heterobilayer in the MBZ. We note a doubly-degenerate valence band maximum (VBM) at the $\Gamma_M$ point and a four-fold degenerate band below it; the conduction band minimum (CBM) is at the $\Lambda_M$ and $\Sigma_M$ points. 
We unfold the wavefunctions from the MBZ to the UBZ of the individual MoSe$_2$ and MoS$_2$ layers.
$\mathbf{k}_M$, a k-point of the MBZ, maps to a k-point, $\mathbf{k}$, restricted to be within the UBZ of layer $l$, through $\mathbf{k}_M$ + $\mathbf{G}^l$
= $\mathbf{k}$, where $\mathbf{G}^l$ is one of 19 reciprocal lattice vectors (RLVs) of the moir\'e lattice for the examined case (Fig. \ref{fig:fig1}(b)). 
For a given $\mathbf{k}$, this map determines unique values of $\mathbf{k}_M$ and $\mathbf{G}^l$, which we denote as $\mathbf{k}_M(\mathbf{k})$ and $\mathbf{G}^l_{\mathbf{k}}$, respectively~\cite{unfold}.
A moir\'{e} eigenstate of the heterostructure ($\psi_{n_M{\textbf{k}_M}}(\textbf{r})$) can be expressed in terms of the individual layer unit cell eigenstates. To identify the layer contribution, we separate the moir\'{e} superlattice eigenstates into two parts along the out-of-plane direction $z$ separating the layers:
\begin{equation}
    \psi_{n_M\textbf{k}_M}(\textbf{r}) = \tilde{\psi}^{Se}_{n_M{\textbf{k}_M}}(\textbf{r}) + \tilde{\psi}^S_{n_M{\textbf{k}_M}}(\textbf{r})
\end{equation}
where $\tilde{\psi}^{Se}_{n_M{\textbf{k}_M}}(\textbf{r})$ contains the wavefunction contribution from the MoSe$_2$ layer via $\tilde{\psi}^{Se}_{n_M{\textbf{k}_M}}(\textbf{r}) = \psi_{n_M{\textbf{k}_M}}(\textbf{r})\Theta(0.5-z)$ and $\tilde{\psi}^{S}_{n_M{\textbf{k}_M}}(\textbf{r}) = \psi_{n_M{\textbf{k}_M}}(\textbf{r})\Theta(z-0.5)$.  
$\Theta$ is the Heaviside step function, and the heterostructure is placed with its mean position at 0.5 in crystal units along $z$, the coordinate in the out-of-plane direction.

We expand the $\tilde{\psi}$'s in terms of the unit-cell eigenstates of layer ($l$) ($\phi_{n\textbf{k}}^l(\textbf{r})$):
\begin{equation}
\tilde{\psi}_{n_M{\textbf{k}_M}}^l(\textbf{r})=\sum_{n\textbf{G}^l}F^{l}_{n\textbf{G}^l,n_M\textbf{k}_M}\phi_{n\textbf{k}_M+\textbf{G}^l}^l(\textbf{r})
    \label{tab:eqn1}
\end{equation}
where $F^{l}_{n\textbf{G}^l,n_M\textbf{k}_M}$, the expansion coefficient, vanishes if $\textbf{G}^l + \textbf{k}_M$ falls outside the UBZ of  layer $l$. Hence for every $\textbf{k}_M$  it is nonzero only for 12 (for MoSe$_2$) or 13 (for MoS$_2$)  of the 19 RLV's in Fig. 1(b). 
Finally, the spectral weight is defined by summing over bands in the unit cell of $l$th layer \cite{unfold},
\begin{equation}
    P^l(n_M,
    \mathbf{k}) = \sum_{n}\lvert F^{l}_{n\textbf{G}^l_{\textbf{k}},n_M\textbf{k}_M(\mathbf{k})}\lvert ^2
    \label{tab:eqn2}
\end{equation}
denoting the probabilistic contribution of the unit cell eigenstates to the moir\'{e} superlattice eigenstate. 

Fig.s \ref{fig:fig1}(d) and (e) show the unfolded spectral weight $P$ for the MoSe$_2$ and MoS$_2$ layers, respectively, plotted against $\mathbf{k}$ and the band energies $\epsilon_{n_M\mathbf{k}_M(\mathbf{k})}$ along specific paths in the UBZ of the respective monolayers. Hole and electron contributions are represented in red and blue colorbars, respectively. As expected, we find that the VBM of the moir\'{e} superlattice arises from the K$_{Se}$ point of MoSe$_2$. The lower-energy four-fold degenerate states at $\Gamma_M$ originate from the $\Gamma$ points of both the layers and from the spin-split K$_{Se}$. This degeneracy is  accidental and specific to the examined twist angle. The valence band edge wavefunctions of both layers at the $\Gamma$ point of the UBZ are delocalized along the out-of-plane direction and hence hybridize substantially in the heterostrutcure \cite{layerdep}. Consequently, the corresponding energies are higher compared to their monolayer bands (shown in grey lines in fig. \ref{fig:fig1}(d) and (e)), unlike the K-localized bands in which the energies remain similar for the monolayer and the heterostructure at the DFT level, which excludes non-local screening. 
The K valley band edge of MoS$_2$ is lower in energy than that of MoSe$_2$, showing clearly the type-II nature of this heterostructure. 
The CBM of the moir\'{e} superlattice originates from the $\Lambda_{S}$ point. Notably, due to interlayer hybridization at this k-point, the CBM shows contribution from $\Lambda_{Se}$ as well. This contribution is absent in the separated-monolayer bandstructures, and as we will show below, dictates the nature of the low-energy excitons in this system. 

Next, we compute the exciton states in the examined heterostructure using GW-BSE. 
We evaluate the dielectric screening and quasiparticle self-energy corrections from G$_0$W$_0$~\cite{BGW}, 
within the generalized plasmon-pole approximation~\cite{gpp} and using spinor wavefunctions (see SI for full computational details). We note that these computations are highly cumbersome and can be achieved owing to an advanced accelerated large-scale version of the BerkeleyGW code~\cite{del2020accelerating, del2019large}.  
The resulting GW interlayer bandgap is 1.74 eV, compared to the DFT bandgap of 1.16 eV. The GW direct intralayer gaps  are 2.46 eV for MoSe$_2$ and 2.97 for MoS$_2$. These GW gaps are somewhat larger than the corresponding monolayer gaps of $\sim$2.3~eV for MoSe$_2$~\cite{refaely2018defect, ugeda2014giant} and $\sim$2.6~eV MoS$_2$~\cite{qiu2016screening}. We associate this gap increase to the band hybridization taking place within the intralayer MoSe$_2$ conduction region and MoS$_2$ valence region, both of which are deep inside the heterostructure band manifolds.

\begin{figure}
    \centering
    \includegraphics[scale=0.25]{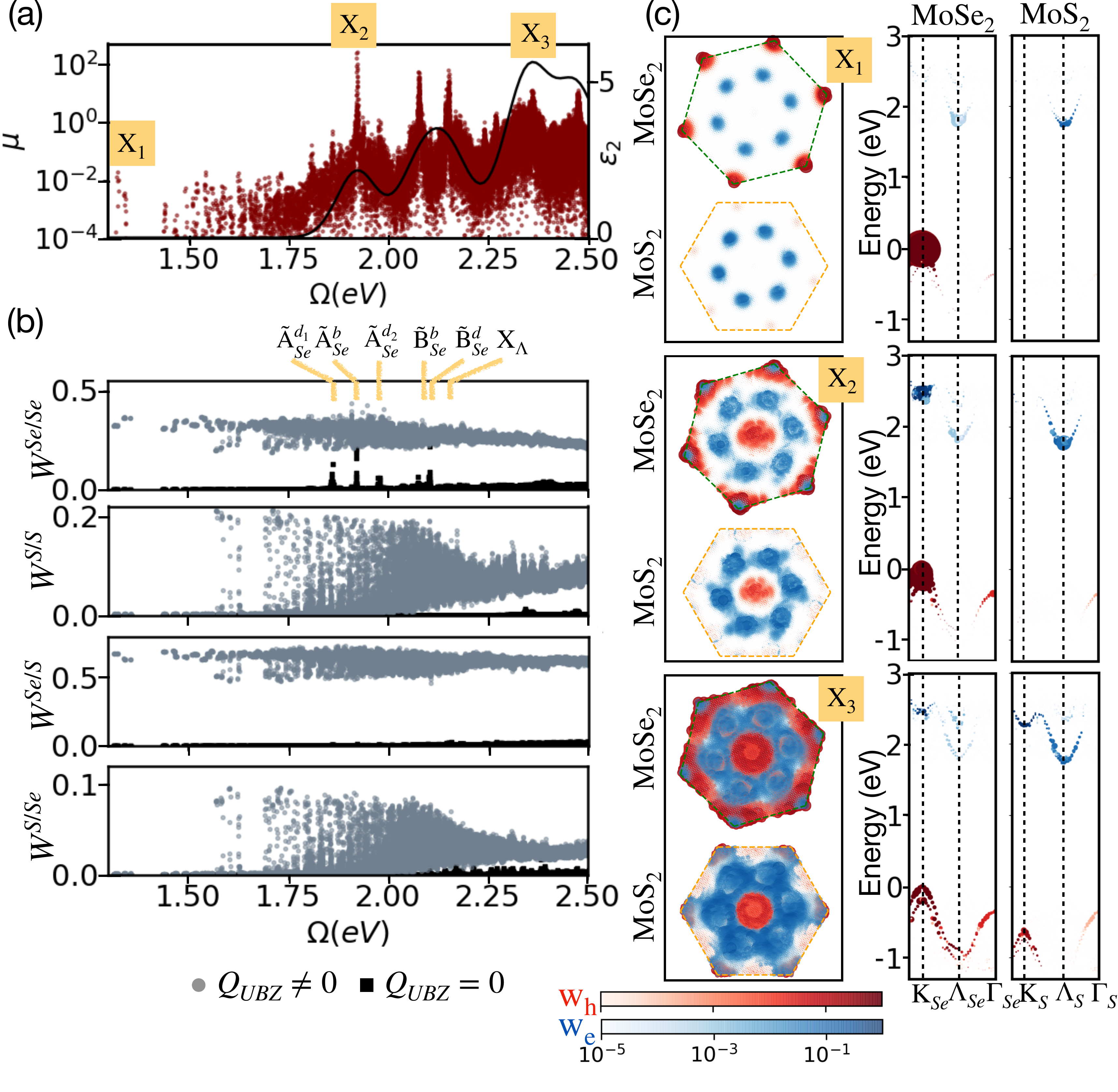}
    \caption{(a): Dipole matrix elements ($\mu$) (on a semi-log scale) and the absorption spectrum ($\varepsilon_2$) of the heterostructure are shown as maroon dots and a black line, respectively. (b) Layer-resolved
    contributions to the excitons as a function of excitation energy. The y-axis shows the contributions associated with the various interlayer and intralayer components. Black squares/grey dots represent contributions from direct/indirect excitons in the UBZ. (c) Momentum and layer resolved electron and hole contributions to the excitons marked as $\textrm{X}_1$, $\textrm{X}_2$ and $\textrm{X}_3$ in the UBZs of MoSe$_2$ and MoS$_2$, as well as the associated band contributions, computed using GW-BSE within the exciton unfolding scheme described in the text. $\textrm{w}_h$ and $\textrm{w}_e$ are the magnitudes of the hole and electron contributions in the UBZ.
    }
    \label{fig:fig2}
\end{figure}

We further solve the BSE equation for the moir\'e system. The $X$th exciton wavefunction ($\Psi^X$) is expressed in the electron-hole basis as:
\begin{equation}
\Psi^X(\textbf{r},\textbf{r}^{\prime}) = \sum_{v_M c_M \textbf{k}_M}A^X_{v_Mc_M\textbf{k}_M}\psi_{c_M\textbf{k}_M}(\textbf{r})\psi^{*}_{v_M\textbf{k}_M}(\textbf{r}^{\prime})
\label{tab:eqn4}
\end{equation}
Here, $A$ are the exciton spanning coefficients and $v_M$,$c_M$ are the valence (hole) and conduction (electron) moir\'e bands, respectively. 
Fig.~\ref{fig:fig2}(a) shows the computed GW-BSE absorption spectrum ($\varepsilon_2$) (black line) and the corresponding electron-hole transition dipole matrix elements ($\mu$) (maroon dots) for light polarized along the $\hat{a}$ in-plane lattice direction, as a function of the exciton energy $\Omega$. 
Notably, our GW-BSE calculation results in a large number of exciton states at the low energy regime, manifesting that there are multiple allowed band-to-band optical transitions induced by the bilayer composition.

To track the origins of this complex absorption structure, we further unfold the computed GW-BSE excitons.
We define a measure of the contributions to the excitons arising from holes at layer $l_2$ and electrons at layer $l_1$, via:
\begin{equation}
\begin{split}
    W^{l_1/l_2}_{X}=&
    \sum_{\mathbf{k}\mathbf{k}^{\prime}}\sum_{v_Mc_M}P^{l_1}(c_M,
    \textbf{k})P^{l_2}(v_M,
    \textbf{k}^{\prime})\\
     &\times \lvert A^X_{v_Mc_M\textbf{k}_M(\textbf{k})}\lvert^2 \, \delta_{\textbf{k}_M(\textbf{k}^{\prime}),\textbf{k}_M(\textbf{k})}
    \label{tab:eqn5}
\end{split}
\end{equation}
where $l_1,l_2$ represent the MoSe$_2$ and MoS$_2$ monolayers (see SI for full details). The $\delta_{\textbf{k}_M(\textbf{k}^{\prime}),\textbf{k}_M(\textbf{k})}$ ensures that the transitions are direct in the MBZ. However, as multiple $\mathbf{k}$ points map to the same $\mathbf{k}_M$, indirect transitions in UBZs are allowed.
Figure~\ref{fig:fig2}(b) shows the resulting inter- and intra-layer exciton components and their momentum directness in the UBZ. Grey dots / black squares represent contributions from indirect ($\textbf{k}\neq\textbf{k}^{\prime}$) / direct ($\textbf{k}=\textbf{k}^{\prime}$) excitons in the UBZ ($Q_{UBZ}\neq0$ / $Q_{UBZ}=0$). Notably, all the computed exciton states mostly originate from  electron-hole transitions that are indirect in the UBZ. 
In addition, specific absorption peaks have significant contribution from intralayer transitions that are direct in the MBZ.   

To understand these results, we further analyze the UBZ momentum- (and band-resolved) contributions. We unfold the electron contribution of the $X$th exciton to the UBZ of layer $l$ via
\begin{equation}
    \textrm{w}^l_{e}(\textbf{k}) =\sum_{c_M} P^l(c_M,
    \textbf{k})\sum_{v_M}\lvert A^X_{v_Mc_M\textbf{k}_M(\textbf{k})}\lvert^2 
    \label{tab:eqn6}
\end{equation}
and similarly for the hole contribution, 
\begin{equation}
    \textrm{w}^l_{h}(\textbf{k}) =\sum_{v_M} P^l(v_M,
    \textbf{k})\sum_{c_M}\lvert A^X_{v_Mc_M\textbf{k}_M(\textbf{k})}\lvert^2
    \label{tab:eqn7}
\end{equation}
Fig.~\ref{fig:fig2}(c) shows the unfolded contributions in the UBZs of MoSe$_2$ and MoS$_2$ for the three excitonic regions marked on the absorption spectrum of Fig.~\ref{fig:fig2}(a): $\textrm{X}_1$, the lowest-energy exciton, which has only indirect UBZ momentum contributions; and $\textrm{X}_2$, $\textrm{X}_3$ - the two lowest excitons that have significant intralayer and direct UBZ momentum contributions within the MoSe$_2$ and MoS$_2$ layers, respectively.  

We first note that while the holes of $\textrm{X}_1$ are localized at the K$_{Se}$ and K$^{\prime}_{Se}$ valleys of the MoSe$_2$ layer, 
due to wavefunction hybridization at the $\Lambda$ valley on both layers, the electron contribution is from both $\Lambda_{Se}$ of MoSe$_2$ and $\Lambda_S$ of MoS$_2$. Thus, the low-energy excitonic region $\textrm{X}_1$, while primarily of interlayer nature, also contains a large intralayer component due to the $\Lambda$ valley contribution. Furthermore, the electrons and holes arise from two different momentum points, namely, these excitons are optically allowed, but  \textit{momentum indirect in the UBZ} ($Q_{UBZ}\neq0$). We emphasize that such transitions become allowed due to the moir\'{e} potential, including both the relative rotation between the layers as well as the atomic reconstruction. The band-resolved contributions, shown along the computed GW bands in Fig.~\ref{fig:fig2}(c), further emphasizes this mixed excitonic nature. 

In the case of $\textrm{X}_2$, both the holes and electrons composing the excitons have finite contributions from the K$_{Se}$ and K$^{\prime}_{Se}$ valleys of the MoSe$_2$ layer. This implies that $\textrm{X}_2$ has an intralayer, momentum-direct component in the UBZ ($Q_{UBZ}=0$) of MoSe$_2$. However, these excitons also consist of a large number of $Q_{UBZ}\neq0$ intralyer and interlayer transitions, mainly coupling holes at the lower spin-split valence band at K$^{\prime}_{Se}$ and electrons at $\Lambda$. $\textrm{X}_3$ is the lowest excitonic state which exhibits intralayer $Q_{UBZ}=0$ components from the MoS$_2$ layer, in addition to higher-energy interlayer transitions and MoSe$_2$ intralayer contributions. Thus, the emergent excitons in the twisted MoS$_2$-MoSe$_2$ bilayer involve both layer-hybridized and momentum-mixed transitions, specifically induced by the twist angle.

\begin{figure}
    \centering
    \includegraphics[scale=0.35]{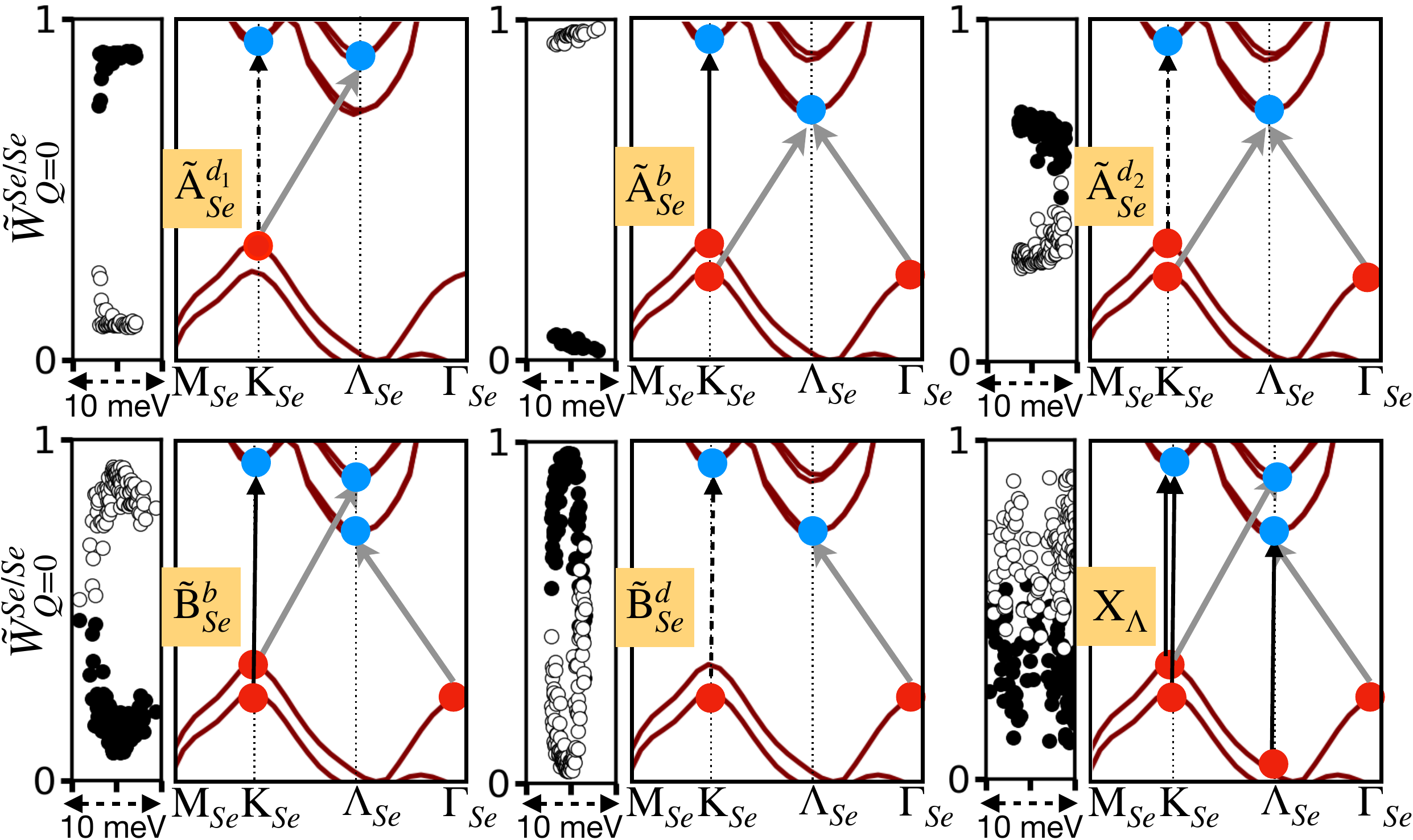}
    \caption{Analysis of the excitonic states with dominant contributions from the MoSe$_2$ intralayer excitations. For each state, the left panels show spin-allowed (white circles) and spin-forbidden (black circles) contributions to the UBZ momentum-direct transitions in an energy range of 10 meV around the excitation energy. The right panels show the unfolded bandstructure of MoSe$_2$, with red and blue dots schematically representing the main hole and electron transitions contributing to each state. Black arrows represent direct and optically bright (solid lines) or dark (dashed lines) K-K transitions in the UBZ, and grey arrows represent the additional momentum indirect contributions in the UBZ.}
    \label{fig:fig3}
\end{figure}

Finally, we analyze the exciton series with significant contributions from intralayer MoSe$_2$ transitions, marked with yellow lines in the upper panel of  Fig.~\ref{fig:fig2}(b). We note that these states appear around 1.85 eV; this energy is higher than the computed A exciton energy in the separate monolayer, of $\sim$1.65-1.75 eV \cite{lu2019interlayer, refaely2018defect, katznelson2022bright}, due to the larger GW quasiparticle gap associated with the hybridized nature of the bands.  Using our unfolding scheme, these states can be directly compared to the well-studied dark and bright excitons composing the low-energy spectrum of the separated monolayer. Fig.~\ref{fig:fig3} shows the band components of these states, along with the spin component of the momentum-direct contribution, labeled as $\bar{A}^{d/b}$ and $\bar{B}^{d/b}$ to connect with the familiar picture associated with the direct UBZ transition at K, and its dark ($d$) and bright ($b$) nature due to spin selection rules. 
The lowest intralayer excitation $\tilde{A}^{d_1}_{Se}$ is dark, due to its dominant spin-forbidden component. Along with direct transition at the $K_{Se}$, it is also composed of a $K_{Se}$ to $\Lambda_{Se}$ transition. At a higher energy we find another dark state, $\tilde{A}^{d_2}_{Se}$, with similar features, but mixed with transitions from the spin-split valence band at $K_{Se}$ to the $\Lambda_{Se}$ conduction band originating from hybridization with the MoS$_2$ layer. The difference in coupling between the hole that is purely on the MoSe$_2$ layer and the $\Lambda$ electrons that are on both layers, but with more contributions from MoSe$_2$ in one and in MoS$_2$ in the second, is responsible for the energy difference between these states.  
The exciton composed of spin-allowed $K$-$K$ transition, $\tilde{A}^{b}_{Se}$, is in between these dark states, and is similar to the $X_2$ state discussed above.  The next two states with large intralayer contributions are the  $\tilde{B}^{b}_{Se}$ and $\tilde{B}^{d}_{Se}$ excitons, as may be expected; however, both include strong additional contributions from holes at the $\Gamma$ point and electrons at $\Lambda$. Surprisingly, we observe another intralayer exciton in this energy region, $X_{\Lambda}$, in which the momentum-direct UBZ contributions are primarily at the $\Lambda$ point. This state is highly unexpected, and results, once again, from the mixed nature of the conduction band at $\Lambda$. 

The change in absorption features compared to the monolayer case is a useful case study for changes arising from twisting and allowing optical transitions which are absent in the case of separate monolayers. The excitons associated with these features are layer delocalized, and cannot be classified as inter/intra-layer states. In particular, these findings challenge the common assumption that large twist angles suppress interlayer coupling due to lattice mismatch~\cite{regan2022emerging}; in fact, our results demonstrate that large twist angles can introduce significant interlayer exciton contributions.

To conclude, we have presented a GW-BSE-based unfolding approach to analyze the absorption spectra and exciton properties as a function of the interlayer twist angle in TMD heterostructures. By including the structural changes due to atomic reconstruction, as well as a momentum mismatch associated with the chosen twist angle, we have shown that electron-hole coupling between different points in the UBZs of the monolayers is not only allowed but can in fact dictate the nature of the excitons. As a result, we find that the exciton fine structure is composed of largely-mixed interlayer and intralayer contributions, which are tunable with the twist angle, and can be expected to change the exciton relaxation dynamics. Our method is general and offers a way to analyze the subtle changes in the optical selection rules arising from the mixing of wavefunctions of different momenta in the UBZ's due to the moir\'{e} potential induced by interlayer twisting.

\textbf{Acknowledgments:} We thank Paulina Plochocka, Keshav Dani, and Ouri Karni for valuable discussions. 
T.A. is supported by the David Lopatie Fellows Program. S.R.A. is an incumbent of the Leah Omenn Career Development Chair and acknowledges support from a Peter and Patricia Gruber Award and an Alon Fellowship, as well as an Israel Science
Foundation Grant No. 1208/19. M.J. and H.R.K. gratefully acknowledge the National Supercomputing Mission of the Department of Science and Technology, India, and the Science and Engineering Research Board of the Department of Science and Technology, India, for financial support under Grants No. DST/NSM/R\&D\_HPC\_Applications/2021/23 and No. SB/DF/005/2017, respectively.
Computational resources were provided by the Oak Ridge Leadership Computing Facility through the Innovative and Novel Computational Impact on Theory and Experiment (INCITE) program, which is a DOE Office of Science User Facility supported under Contract No. DE-AC05-00OR22725; Supercomputer Education and Research Center at Indian Institute of Science; and the ChemFarm cluster at the Weizmann Institute of Science.

\noindent 
Emails of corresponding authors: mjain@iisc.ac.in, sivan.refaely-abramson@weizmann.ac.il

\bibliography{ref}
\end{document}


\author{Sudipta Kundu}
\altaffiliation[Present address: ]{Department of Materials Science and Engineering, Stanford University, Stanford, CA, USA}
\affiliation{Centre for Condensed Matter Theory, Department of Physics, Indian Institute of Science, Bangalore 560012, India}
\author{Tomer Amit}
\affiliation{Department of Molecular
Chemistry and Materials Science, Weizmann Institute of
Science, Rehovot 7610001, Israel}
\author{H. R. Krishnamurthy}
\affiliation{Centre for Condensed Matter Theory, Department of Physics, Indian Institute of Science, Bangalore 560012, India}
\author{Manish Jain$^*$}
\affiliation{Centre for Condensed Matter Theory, Department of Physics, Indian Institute of Science, Bangalore 560012, India}
\author{Sivan Refaely-Abramson$^*$}
\affiliation{Department of Molecular
Chemistry and Materials Science, Weizmann Institute of
Science, Rehovot 7610001, Israel}

\title{Supplementary information for `Exciton fine structure in twisted transition metal dichalcogenide heterostructures'}
\maketitle

\nopagebreak

\section{Computational details}

The MoS$_2$/MoSe$_2$ heterostructure with 16\si{\degree} interlayer twist angle is constructed using the Twister package \cite{naiktwister2}.
The mean-field electronic structures are calculated using density functional theory as implemented in Quantum Espresso package \cite{qe}. We use optimized norm-conserving Vanderbilt pseudopotentials \cite{oncv}. The exchange-correlation functional is approximated using the Perdew-Zunger parameterization of the local density approximation \cite{pz}. The wavefunctions are expanded in a plane-wave basis with plane-wave energies up to 60 Ry. The moir\'e Brillouin zone (MBZ) is sampled with 3$\times$3$\times$1 Monkhorst-Pack grid of k-points \cite{kpt}. We further employ van der Waals corrections
using the van der Waals corrected density functional along with Cooper exchange \cite{dfc09} to relax the twisted heterostructure.  The structure is relaxed until the force on every atom is less than 25 meV/\AA{}. For the electronic structure calculations, we use fully relativistic pseudopotentials including explicit spin-orbit coupling. As the van der Waals functional does not
affect the bandstructure substantially, we do not use the van der Waals correction functional in the bandstructure calculation.

We compute the exciton states in the examined heterostructure using GW-BSE \cite{BGW}. We evaluate the dielectric screening and quasiparticle self-energy corrections from G$_0$W$_0$~\cite{gpp}, using a  3$\times$3$\times$1 k-points grid with an additional subsampled non-uniform grid \cite{felipe2017nonuniform}. The calculation of  dielectric function includes 5000 non-spinor electronic bands (325 of them occupied). The dielectric matrix is expanded in plane waves with energy up to 25 Ry.
We use a truncation scheme for the Coulomb interaction between the periodic layers  along the $z$ axis. We compute the self-energy within the generalized plasmon-pole approximation \cite{gpp} and with spinor wavefunctions, including 2000 bands. The static
remainder technique is employed to accelerate the convergence with the number of bands \cite{static_ch}.
We include 30 occupied and 30 unoccupied spinor wavefunctions in the BSE Hamiltonian. We use a 6$\times$6$\times$1 coarse grid sampling of the MBZ, interpolated on a finer 18$\times$18$\times$1 grid to compute the matrix element of the electron-hole interaction kernel.  
We note that this grid for the chosen supercell is equivalent to a $\sim$76$\times$76$\times$1 grid in the unit cell. It is also to be noted that we use an advanced accelerated GPU-based large-scale version of the BerkeleyGW code \cite{del2019large,del2020accelerating}.

\section{Formalism for exciton unfolding}
An exciton wave function is written in the electron-hole basis as:
        \begin{equation}
        \label{eqns1}
        \Psi^X(\textbf{r},\textbf{r}^{\prime}) = \sum_{v_Mc_M\textbf{k}_M}A^X_{v_Mc_M\textbf{k}_M}\psi_{c_M\textbf{k}_M}(\textbf{r})\psi^{*}_{v_M\textbf{k}_M}(\textbf{r}^{\prime})
        \end{equation}
        Dividing $\psi_{c\textbf{k}_M}$ into two parts belonging to each of the layers, Eqn. 
        \ref{eqns1} becomes:
        \begin{equation}
            \label{eqns2}
            \Psi^X(\textbf{r},\textbf{r}^{\prime}) = \sum_{v_Mc_M\textbf{k}_Ml_1l_2}A^X_{v_Mc_M\textbf{k}_M}\psi^{l_1}_{c_M\textbf{k}_M}(\textbf{r})\psi^{*l_2}_{v_M\textbf{k}_M}(\textbf{r}^{\prime})
        \end{equation}
        where $l_1,l_2$ can correspond to either MoS$_2$ and MoSe$_2$ layers. Intralayer contributions correspond to $l_1=l_2$, and interlayer contributions to $l_1\neq l_2$. Each of the $\textbf{k}_M$ points in MBZ unfolds to 12 and 13 $\textbf{k}$ points in the unit cell Brillouin Zones (UBZs) of MoSe$_2$ and MoS$_2$ layers via 12 and 13 moir\'e reciprocal lattice vectors $\textbf{G}^l$, respectively. To understand the contributions from the momenta in the different UBZs to the excitons, we write the electron-hole basis ($\psi$'s) in terms of the UBZ eigenstates (see Eqn. 2 of the main text) and compute the squared norm of $\Psi^X$. It is straightforward to verify that  this can be expressed as 
        \begin{equation}
            \label{eqns3}
            \lVert \Psi^X\lVert^2 = \sum_{\textbf{k}\textbf{k}^{\prime}}\sum_{v_Mc_Ml_1l_2}\lvert A^X_{v_Mc_M\textbf{k}_M(\textbf{k})}\lvert^2{P}^{l_1}(c_M,\textbf{k}) {P}^{l_2}(v_M,\textbf{k}^{\prime})\delta_{\textbf{k}_M(\textbf{k}^{\prime}),\textbf{k}_M(\textbf{k})}
        \end{equation}
        where the functions $P^l$s are as defined in Eqn. 3 of the main text. 
        Depending on the values of $l_1$ and $l_2$, there are four terms contributing to $\lVert\Psi^X\lVert^2$.\\
        \textbf{Case 1}: $l_1,l_2$ both correspond to MoSe$_2$ layer.\\
        $W^{Se/Se}_X=\sum_{\textbf{k},\textbf{k}^{\prime}}\sum_{v_Mc_M}{P}^{Se}{(c_M,\textbf{k})} P^{Se}{(v_M,\textbf{k}^{\prime})}\lvert A^X_{v_Mc_M\textbf{k}_M(\textbf{k})}\lvert^2 \, \delta_{\textbf{k}_M(\textbf{k}^{\prime}),\textbf{k}_M(\textbf{k})}$
        \\
        \textbf{Case 2}: $l_1,l_2$ both correspond to MoS$_2$ layer.\\
        $W^{S/S}_X=\sum_{\textbf{k},\textbf{k}^{\prime}}\sum_{v_Mc_M}{P}^{S}{(c_M,\textbf{k})} P^{S}{(v_M,\textbf{k}^{\prime})}\lvert A^X_{v_Mc_M\textbf{k}_M(\textbf{k})}\lvert^2 \, \delta_{\textbf{k}_M(\textbf{k}^{\prime}),\textbf{k}_M(\textbf{k})}$
        \\
        \textbf{Case 3}: $l_1,l_2$ correspond to MoS$_2$ and MoSe$_2$ layer respectively.\\
        $W^{S/Se}_X=\sum_{\textbf{k},\textbf{k}^{\prime}}\sum_{v_Mc_M}P^{S}{(c_M,\textbf{k})} P^{Se}(v_M,\textbf{k}^{\prime})\lvert A^X_{v_Mc_M\textbf{k}_M(\textbf{k})}\lvert^2 \, \delta_{\textbf{k}_M(\textbf{k}^{\prime}),\textbf{k}_M(\textbf{k})}$
        \\
        \textbf{Case 4}: $l_1,l_2$ correspond to MoSe$_2$ and MoS$_2$ layer respectively.\\
        $W^{Se/S}_X=\sum_{\textbf{k},\textbf{k}^{\prime}}\sum_{v_Mc_M}P^{Se}{(c_M,\textbf{k})} P^{S}(v_M,\textbf{k}^{\prime})\lvert A^X_{v_Mc_M\textbf{k}_M(\textbf{k})}\lvert^2 \, \delta_{\textbf{k}_M(\textbf{k}^{\prime}),\textbf{k}_M(\textbf{k})}$
        \\
        In each of these 4 cases, terms with $\textbf{k}=\textbf{k}^{\prime}$ correspond to contributions from the direct transitions in the UBZs and the rest to the indirect transitions in the UBZs. These computed contributions are shown in Fig. 2(b) in the main text.

\section{Spin contributions}
To identify the excitons with large intralayer contributions from MoSe$_2$, we extend our unfolding scheme to find the dominant spin contribution to the momentum-direct components in the UBZ.
In the eigenstate basis of $\sigma_z$ , the electron and hole wavefunctions are written as:
                \begin{align}
                \label{eqns4}
                \psi_{c_M}^{l_1} &= \begin{bmatrix}
                \psi_{c_M\uparrow}^{l_1} \\
                \psi_{c_M\downarrow}^{l_1}
                                \end{bmatrix}
                \end{align}
                and
                \begin{align}
                \label{eqns5}
                \psi_{v_M}^{l_2} &= \begin{bmatrix}
                \psi_{v_M\uparrow}^{l_2} \\
                \psi_{v_M\downarrow}^{l_2}
                \end{bmatrix}
                \end{align}
                The electron-hole basis becomes:
                \begin{align}
                \label{eqns6}
                \psi_{c_M}^{l_1}\otimes\psi^{*l_2}_{v_M} &= \begin{bmatrix}
                \psi_{c_M\uparrow}^{l_1}\psi^{*l_2}_{v_M\uparrow} \\
                \psi_{c_M\uparrow}^{l_1}\psi^{*l_2}_{v_M\downarrow}\\
                \psi_{c_M\downarrow}^{l_1}\psi^{*l_2}_{v_M\uparrow} \\
                \psi_{c_M\downarrow}^{l_1}\psi^{*l_2}_{v_M\downarrow}
                \end{bmatrix}
                \end{align}
The squared norm of the exciton eigenstate can then be rewritten as:
\begin{equation}
\label{eqns7}
                \lVert\Psi^X\lVert^2 = \sum_{v_Mc_M\textbf{k}_Ml_1l_2}\lvert A^X_{v_Mc_M\textbf{k}_M}\lvert^2(\lVert\psi_{c_M\uparrow}\psi^*_{v_M\uparrow}\lVert^2+\lVert\psi_{c_M\uparrow}\psi^*_{v_M\downarrow}\lVert^2+\lVert\psi_{c_M\downarrow}\psi^*_{v_M\uparrow}\lVert^2+\lVert\psi_{c_M\downarrow}\psi^*_{v_M\downarrow}\lVert^2)
                \end{equation}
                Following the same notation as defined in the main text for $P^l$'s but accounting for the spinor parts of the wavefunctions:
                \begin{align}
                \label{eqns8}
                \lVert\Psi^X\lVert^2 & =  \sum_{\textbf{k}\textbf{k}^{\prime}}\sum_{v_Mc_Ml_1l_2} [{P}^{l_1\uparrow}(c_M,\textbf{k}){P}^{l_2\uparrow}(v_M,\textbf{k}^{\prime})+{P}^{l_1\downarrow}(c_M,\textbf{k}){P}^{l_2\uparrow}(v_M,\textbf{k}^{\prime})+\\ \nonumber
                & {P}^{l_1\uparrow}(c_M,\textbf{k}){P}^{l_2\downarrow}(v_M,\textbf{k}^{\prime})+{P}^{l_1,\downarrow}(c_M,\textbf{k}){P}^{l_2\downarrow}(v_M,\textbf{k}^{\prime})]\\ \nonumber
                & \times \lvert A^X_{v_Mc_M\textbf{k}_M(\textbf{k})}\lvert^2 \, \delta_{\textbf{k}_M(\textbf{k}^{\prime}),\textbf{k}_M(\textbf{k})}\\ \nonumber
                & = \sum_{l_1,l_2}\large[\tilde{W}^{l_1/l_2}_{spin-forbidden}+\tilde{W}^{l_1/l_2}_{spin-allowed}\large]
                \end{align}
                For a particular choice of $l_1,l_2$, the spin-allowed (like spin of conduction and valence wavefunction) contribution is given by:
                \begin{align}
                \label{eqns9}
                    \tilde{W}^{l_1/l_2}_{spin-allowed} = & \sum_{\textbf{k}\textbf{k}^{\prime}}\sum_{v_Mc_M}[{P}^{l_1\uparrow}(c_M,\textbf{k}){P}^{l_2\uparrow}(v_M,\textbf{k}^{\prime})+{P}^{l_1,\downarrow}(c_M,\textbf{k}){P}^{l_2\downarrow}(v_M,\textbf{k}^{\prime})]\ \\ \nonumber
                    & \times \lvert A^X_{v_Mc_M\textbf{k}_M(\textbf{k})}\lvert^2 \, \delta_{\textbf{k}_M(\textbf{k}^{\prime}),\textbf{k}_M(\textbf{k})}
                \end{align}
                \begin{align}
                \label{eqns10}
                    \tilde{W}^{l_1/l_2}_{spin-forbidden} = & \sum_{\textbf{k}\textbf{k}^{\prime}}\sum_{v_Mc_M}[{P}^{l_1\uparrow}(c_M,\textbf{k}){P}^{l_2\downarrow}(v_M,\textbf{k}^{\prime})+{P}^{l_1,\downarrow}(c_M,\textbf{k}){P}^{l_2\uparrow}(v_M,\textbf{k}^{\prime})]\ \\ \nonumber
                    & \times \lvert A^X_{v_Mc_M\textbf{k}_M(\textbf{k})}\lvert^2 \, \delta_{\textbf{k}_M(\textbf{k}^{\prime}),\textbf{k}_M(\textbf{k})}
                \end{align}
The $\textbf{k}=\textbf{k}^{\prime}$ terms in Eqn. \ref{eqns9} and \ref{eqns10} give the spin-allowed and forbidden contributions to the direct transitions in UBZ. For $l_1=l_2$=MoSe$_2$, the spin-allowed and spin-forbidden contributions to the direct transitions in the MoSe$_2$ UBZ are depicted in Fig. 3 of the main text by the white and black circles, respectively.

\bibliography{ref}